\documentclass[aps,prb,twocolumn,superscriptaddress,showpacs]{revtex4-2}
\usepackage{bm}
\usepackage{amsmath}
\usepackage{amssymb}
\usepackage{graphicx}
\graphicspath{{img/}}  
\usepackage{color}

\begin{document}

\title{Thermoelectric effects in tunneling of spin-polarized electrons in a molecular transistor}

\author{A.D. Shkop}
\affiliation{B. Verkin Institute for Low Temperature Physics and
Engineering of the National Academy of Sciences of Ukraine, 47
Prospekt Nauky, Kharkiv 61103, Ukraine}
\begin{abstract}
Thermal transmission in a molecular transistor with fully spin-polarized electrodes subjected to a temperature gradient is considered. The problem has been solved by using density matrix method in perturbation approach over small tunneling width. It has been found that due to the vibronic effects spintronic molecular transistor is characterized by negative differential thermoconductance. It has been demonstrated that in the dependence of thermopower $S$ on the detuning energy there is an increased number of points of change of the sign and magnitude of $S$ comparing with that of conventional molecular transistor. Optimal parameters, that provide the highest thermoelectric power at maximum efficiency $P_{me}$ for spintronic molecular transistor, have been found. The dependences of the figure of merit $ZT$ and $P_{me}$ on temperature and an external magnetic field have been calculated and the influence of Coulomb interaction on the thermolelectric properties has been studied. It has been revealed, that for non-zero Coulomb interaction more handy regime of thermoelectric device develops, characterized by continuous region of external magnetic fields, which provide high values of thermoelectric power. 
\end{abstract}

\maketitle

\section*{Introduction}
The scientific community is interested in thermoelectric phenomena due to the possibility of creating thermoelectric devices that convert heat dissipated in physical processes into useful electrical energy (\cite{Benenti},\cite{Di_Ventra},\cite{ReddyPerspective}). Such devices as thermoelectric machines, local coolers, thermal transistors, etc.~are studied in the modern scientific world. Conditions of high efficiency of nanoscale thermoelectric machines are being sought. Quantum dot (QD)-based transistors have high thermoelectric efficiency and good opportunities for integration into microcircuits (\cite{Finch},\cite{FlensbergThermo},\cite{Murphy}).

The theoretical description of thermoelectric phenomena in QD-based nanodevices has been carried out in a number of modern works. An important thermoelectric property of nanoscopic systems is that the sign and magnitude of the thermopower (and the magnitude of other thermoelectric coefficients) depend (\cite{Staring1}) on the detuning energy of the QD level from the Fermi energy. Thermo machines based on gated nanodevices seemed to be efficient (\cite{Murphy},\cite{Kennes},\cite{SierraSanchez},\cite{Swirkowicz},\cite{Parafilo},\cite{Zubov}, see also the reviews \cite{Benenti},\cite{Di_Ventra}) due to known increase of the thermopower and a deviation of the Wiedemann-Franz law $K/(GT)<L_0$ (where $G$ and $K$ are the electric and thermal conductances, $T$ is the average temperature, $L_0$ is the Lorentz number) caused by the Coulomb blockade effect. It was found in Ref.~\cite{SierraSanchez} that the thermoelectric coefficient shows a characteristic Coulomb butterfly structure due to charging effects. In Ref.~\cite{Swirkowicz} magnetic and thermoelectric properties in the transport of partially spin-polarized electrons through a QD in the Coulomb blockade regime were studied.
For heat current in a spin-boson nanojunction model (Ref. \cite{SegalRectifier}) the possibility of occurrence of the phenomenon of rectification when the junction is asymmetric over the couplings to the heat reservoirs was found. In Ref. \cite{SierraSanchez_Proceedings} both voltage and temperature biases for the heat current are considered and non-monotonic dependence of differential thermoconductance on temperature difference due to Coulomb interaction is found for immovable QD. 

Together with conventional single-electron transistors (SETs), single-molecule transistors (SMTs) (\cite{Galperin}) are promising. When studiyng transport properties of SMTs, vibrational degrees of freedom of the molecule must be taken into account. Note that in such a system electron shuttling regime with QD-oscillations of high (classical) amplitude $a$ ($a\gg x_0$, where $x_0$ is the amplitude of zero-point oscillations) can occur (\cite{Shuttle}). Furthermore, small \textit{quantum} oscillations ($a\simeq x_0$) also affect the system. Transport properties at low temperatures are influenced by electron-vibron interaction  (\cite{Flensberg},\cite{Koch}). It has been repeatedly observed experimentally (e.g. \cite{ParkNature},\cite{Monthioux},\cite{Martinez-Blanco}). 

In a molecular transistor, along with the possibility of elastic tunneling there is a possibility of inelastic electron tunneling with emission or absorption of vibrons, i.e. dot vibrations (\cite{Glazman},\cite{Mitra},\cite{Flensberg},\cite{Koch},\cite{Ferone2}). It has been shown that current-voltage characteristics of a molecular transistor at low temperatures has a form of step-like functions with the so-called Frank-Condon steps (\cite{Flensberg},\cite{Koch}). For strong electron-vibron interaction the current through the system at low biases is strongly suppressed (the Frank-Condon blockade).

Thermoelectric properties of the systems with a pronounced electron-vibron interaction have been intensively studied by an increasing number of researchers (\cite{ReddyPerspective},\cite{Khedri},\cite{Tagani2013}, see also the review \cite{Di_Ventra}). In this work the heat current and differential thermoconductance for arbitrary large temperature difference have been studied and, on the other hand, the thermopower, figure of merit $ZT$ and thermoelectric power at maximal efficiency in linear response have been investigated. In the context of this paper, noteworthy is the work \cite{SegalNDTC}, where the phenomenon of negative differential thermoconductance was found in two-level spin-boson model comprising nonlinear interactions. 

A non-monotonic dependence of the thermopower on dot level energy in the systems based on oscillating molecules was studied in Refs. (\cite{KochThermo},\cite{Zianni}). As well as in QD-based devices, in SMTs the dependence of the thermopower on the energy in the molecule changes the sign at the points of resonance transmission and at the points of the electron-hole symmetry of the system. Furthermore, the dependence of the thermopower on the molecule's energy level acquires additional features, namely, the sawtooth behaviour with a step of the order of molecule oscillation quantum, because the channels corresponded to inelastic tunneling processes are added (\cite{KochThermo},\cite{Zianni}). Measurements of the thermopower in molecular junctions have been carried out in a number of modern challenging experiments (e.g., \cite{ReddyExpr2007},\cite{ReddyExpr2019},\cite{Segalman}).

It is interesting to consider the SMT with magnetic electrodes. This makes it possible to control the current by means of an external magnetic field and study a number of spintronic phenomena. Note that the effect of an external magnetic field also manifests itself in changes in the transport characteristics in the presence of magneto-elastic coupling, as shown for suspended carbon nanotubes in Ref. \cite{RastelliEurophys},\cite{RastelliComptes}. It was demonstrated in \cite{PistolesiShekhter} that the coupling between the spin-carrying and polaronic subsystems results in the appearance of localized quantum states. In \cite{KulinichSkorobagatko}, the expressions for the charge current both in the shuttling regime and the regime of low-amplitude quantum oscillations were obtained when taking into account the influence of an external magnetic field on tunneling amplitude phase.

In the systems, which consider double QD, the features of behaviour of thermopower and $ZT$ depending on detuning energy in presence of several tunneling channels were found in, e.g., \cite{Barnas}, \cite{Tagani2012}, \cite{Chi}, \cite{Dai}.

For nanotransistors with partially spin-polarized electrons the conditions for the cooling of the mechanical mode taking into account spin-vibrational interaction and Coulomb repulsion were found (\cite{Stadler2014},\cite{Bruggemann}). Thermoelectric effects in transport of partially spin-polarized electrons through an immovable QD in the absence of an external magnetic field were considered in \cite{Swirkowicz},\cite{Krawiec}, where thermoelectric coefficients and efficiency of the thermoelectric device were found in linear response when accounting for the Coulomb interaction. The possibility to control the development of shuttling by an external magnetic field in the case of partially spin-polarized carriers has been investigated in \cite{Ilinskaya1}.

In this work, the limiting case of fully and oppositely spin-polarized half-metallic electrodes is considered. These materials were predicted in \cite{Groot} and discovered experimentally \cite{Wada}. When the carriers are fully spin-polarized, an external magnetic field directed perpendicular to the leads' magnetizations promotes spin-flips in the dot and enables the tunneling current. For such systems some promising properties were theoretically predicted. Refs.~\cite{Kalinenko}, \cite{Spintromech} studied the conditions of the existence of magnetic shuttle when accounting for the exchange interaction of the spin of the QD with the magnetization vectors of the electrodes. In Ref. \cite{Zubov} the authors study spintronic transistor based on an immovable QD in a magnetic field and calculate transport and thermoelectric characteristics of the junction, investigating in detail the influence of the magnetic field on the efficiency and power of the device.

In \cite{Interplay}, the previous paper of the author of this paper with coauthors, was demonstrated that the number of Franck-Condon steps in the current-voltage characteristics of spintronic molecular transistor is doubled due to the Zeeman splitting. Furthermore, strong Coulomb correlations lead to the appearance of a region without steps. It was noted that in the system with Zeeman splitting Coulomb blockade is lifted by bias voltage in two stages. These feautures directly affect the number of resonance points on the energy scale for concidered system. 

Moreover, it has been predicted in \cite{Monthioux} (see also \cite{Ferone2},\cite{Interplay}), that in molecular transistor temperature dependence of differential electric conductance at intermediate temperatures, $T \simeq \hbar \omega$ ($T$ is the average temperature, $k_B=1$ is the Boltzmann's constant, $\hbar \omega$ is the oscillation quantum), is anomalously non-monotonic. This can lead to unusual effects when studying thermoelectric phenomena.

\begin{figure}\centering
\includegraphics[scale=0.75]{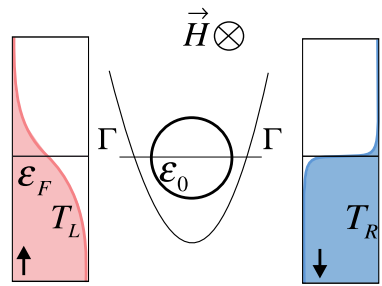}
\caption{Model: a transistor, where vibrating molecule is placed between two magnetic electrodes, fully spin-polarized in opposite directions. Electrodes' temperatures are different, $T_L>T_R$. An external magnetic field $\vec H $ is applied. $\varepsilon_F$ is the Fermi level, $\varepsilon_0$ is the energy of electrons in the molecule, $\Gamma$ is the level width.}
\label{ris:Model_MT_Thermo}
\end{figure}

Following system is considered in this work. A molecule represented as a vibrating QD is placed between two magnetic electrodes, fully spin-polarized in opposite directions (Fig.~\ref{ris:Model_MT_Thermo}). The electrodes are held at different temperatures, $T_L>T_R$. An external magnetic field is applied perpendicular to the leads' magnetizations. The electron-vibron interaction and the Coulomb interaction $U$ in the molecule are taken into account. Tunnel coupling is assumed to be weak (comparing to smallest of elecrodes' temperature), that corresponds to the regime of high thermal efficiency due to sharp transmission resonances (\cite{Murphy}) and common for experimental implementations of molecular transistors. The density matrix of the vibronic subsystem is considered to be in equilibrium, following, for example, \cite{Flensberg},\cite{Koch},\cite{Monthioux},\cite{Ferone2}.

It was found in this work that for non-zero electron-vibron interaction spintronic molecular transistor is characterized by negative differential thermoconductance. The dependence of thermopower on the detuning energy acquires addition points of function's change of the sign and function's inflection, comparing with conventional molecular transistor due to the Zeeman splitting and manifestation of the vibronic effects. Optimal parameters for most effective regime of the device were found, including the dependences of figure of merit $ZT$ and thermoelectric power at maximal efficiency on an external magnetic field. It was shown, that for non-zero Coulomb interaction the device may demonstrate maximal power in continuous range of external magnetic fields, in contrast to the device for $U\rightarrow0$, characterized by two-part region of optimal magnetic fields. All of the features can be used to control the thermoelectric properties by means of an external magnetic field, gate voltage and charging energy, accounting for electron-vibron coupling.

\section{Model and Hamiltonian}
Considered model (Fig.~\ref{ris:Model_MT_Thermo}) is the transistor, where the molecule is placed between two magnetic electrodes, fully spin-polarized in opposite directions. In this problem, the molecule is treated as single-level QD. Its oscillations “as a whole” along the axis parallel to the direction of tunneling current are described as tha oscillations of a harmonic oscillator. The parabola depicts the harmonic potential in which the QD moves. Electrodes’ temperatures are different, $T_L-T_R=\delta T$. An external magnetic field $\vec H$ is applied perpendicular to the plane of the magnetization vectors of the electrodes. For zero magnetic field, there is a “spin blockade” in the system (\cite{Shekhter}). The external magnetic field causes spin-flips in the dot and allows the tunneling current to flow.

The system is characterized by the Hamiltonian 
\begin{equation}
\mathcal{H}= H_l+ H_{d}+H_{v-d}+ H_{tun},
\end{equation}
where $H_l$ is the Hamiltonian of non-interacting electrons in the
leads
\begin{equation}
H_l=\sum_{\alpha}H_{\alpha}=\sum_{k,\alpha}\varepsilon_{k,\alpha}
c^\dag_{k,\alpha}c_{k,\alpha},
\end{equation}
where the operator $c_{k,\alpha}^\dag (c_{k,\alpha})$ creates (annihilates) an electron with the momentum $k$ and the energy $\varepsilon_{k,\alpha}$ in the  $\alpha=L,R$ electrode (full spin polarization allows one to set the lead index $L/R$ coinciding with the spin index $\sigma=\,\uparrow/\downarrow$). The Hamiltonian of electrons in the dot $H_{d}$ reads
\begin{equation}
H_{d}=\sum_\sigma \varepsilon_0 d_\sigma^\dag d_\sigma-\frac{h}{2}(d_\uparrow^\dag d_\downarrow+d_\downarrow^\dag
d_\uparrow)\\
+U d_\uparrow^\dag d_\uparrow d_\downarrow^\dag
d_\downarrow.
\end{equation}
Here the operator $d_\sigma^\dag(d_\sigma)$ is the creation (annihilation)
operator of electron state with spin projection $\sigma$ in the dot. Dot energy level is
$\varepsilon_0$ and it can be tuned by means of gate voltage.
Electron-electron interaction in the dot is denoted by $U$,
$h=g\mu_B H$ is the external magnetic field in energy units, where $g$
is the gyromagnetic ratio, $\mu_B$ is the Bohr magneton.

Vibronic subsystem and electron-vibron interaction are presented via $H_{v-d}$, which reads
\begin{equation}
H_{v-d}=\hbar\omega b^{\dag}b+\varepsilon_{int}
(b^{\dag}+b)(d_\uparrow^\dag d_\uparrow+d_\downarrow^\dag d_\downarrow),
\end{equation}
where the bosonic operator $b^{\dag} (b)$ describes vibronic state in the dot, $\hbar\omega$ is the oscillation quantum, $\varepsilon_{int}$ is the characteristic interaction energy. 

The tunnelling Hamiltonian $H_{tun}$ in a case of full spin polarization of the electrodes reads
\begin{equation}
H_{tun}=t_0\sum _k c^\dag_{k,L}d_\uparrow+t_0\sum _k
c^\dag_{k,R}d_\downarrow+\text{H.c.},
\end{equation}
and tunnel junction is set to be symmetric with the tunneling amplitudes $t_L=t_R=t_0$. The level width of electron state in the dot is proportional to the square of the tunneling amplitude, $\Gamma=2\pi\nu t_0^2$, where $\nu$ is the density of states in the leads, which is assumed to be energy independent (wide-band approximation). 

The Hamiltonian of the system after two unitary transformations: i) the rotation of axes and ii) Lang-Firsov transformation (the description of calculation process has been provided in detail in Ref. \cite{Interplay}) takes the form of a sum of four terms: $H_l$, which remains the same, $H_{d}$, which now reads
\begin{eqnarray}\label{H_dot}
H_{d}=\sum_{s=1}^2 \varepsilon_s a_s^\dag
a_s+(U-2\lambda^2\hbar\omega)a_1^\dag a_1 a_2^\dag a_2,
\end{eqnarray}
where 
$a_1=(1/\sqrt{2})(d_{\uparrow}-d_{\downarrow}),\,a_2=(1/\sqrt{2})(d_{\uparrow}+d_{\downarrow})$ and $\lambda=\varepsilon_{int}/\hbar \omega$ is the electron-vibron interaction constant. It is denoted, that
\begin{equation}\label{E_s}
\varepsilon_{1,2}=\varepsilon_0-\lambda^2\hbar\omega \pm h/2.
\end{equation}
The energy level in the dot is renormalized by electron-vibron interaction. It is also Zeeman split. The strength of the Coulomb interaction is also renormalized due to electron-vibron interaction (note that only the case $U-2\lambda^2\hbar\omega \geqslant 0$ is considered, and the notation $U$ instead of $U-2\lambda^2\hbar\omega$ is used below in the text). The vibronic Hamiltonian $H_{v-d}$ is reduced to
\begin{eqnarray}
H_v=\hbar\omega b^{\dag}b,\label{Hvibr}
\end{eqnarray}
and the tunnel Hamiltonian reads
\begin{equation} \label{8}
H_{tun}=\frac{t_0}{\sqrt{2}}\sum_{k,\alpha}\text{e}^{-\lambda (b^{\dag}-b)}
c^\dag_{k,\alpha}(j_\alpha a_1+a_2)+\text{H.c.},
\end{equation}
where $j_{L,R}=\pm 1$. The $H_{tun}$ acquires an exponential factor, depending on electron-vibron interaction.

The charge current at low temperatures \cite{Interplay} has a form of step-like function, where each current jump is a manifestation of the opening of a new inelastic channel -- tunneling with an emission of vibron.

\section{Thermoelectric properties of spintronic molecular transistor}
Thermoelectric phenomena are considered starting with the heat current. It is found using the standard formula
\begin{equation} 
J_{\alpha}=\text{Tr}\left(\rho\;\dfrac{d}{dt}(H_{\alpha}-\mu_{\alpha} N_{\alpha})\right),
\end{equation}
where $\rho$ is the density matrix of the system, $\mu_{\alpha}$ is the electro-chemical potential of the electrode $\alpha$ and $N_{\alpha}=\sum_{k} c^\dag_{k,\alpha}c_{k,\alpha}$ is the operator of the number of particles. The trace "Tr" is taken over all degrees of freedom of the system. 

The problem is solved using the density matrix method in perturbation theory over small parameter $\Gamma/ \text{min} [T_L, T_R]\ll 1$, i.e.~in the weak tunnel coupling limit. In perturbation approach density matrix of the system is factorized, $\rho\approx \rho_l \otimes \rho_d \otimes \rho_v$, where the density matrices of the leads, dot and vibronic subsystem correspondingly are introduced. The density matrix of vibronic subsystem is assumed to be equilibrium (due to strong vibronic coupling to the environment, so the relaxation rate of the vibronic subsystem is much higher than the tunneling rate). 

The matrix elements of the operator of density matrix of electrons in the dot $\rho_d$ in the space of states in the QD are: $\rho_0=\langle 0\vert\rho_{d}\vert 0\rangle,\;\rho_s=\langle
s\vert\rho_{d}\vert s\rangle,\; s=1,2,\; \rho_D=\langle D\vert\rho_{d}\vert D\rangle,\; \rho_{ss^\prime}=\langle s\vert\rho_{d}\vert s^\prime\rangle,\;s^\prime\neq s =1,2$, where $\vert 0\rangle$ is
the vacuum state, states $\vert s\rangle$ read $\vert s\rangle=a_s^\dag \vert 0\rangle$, and the state $\vert D
\rangle=a_1^\dag a_2^\dag \vert 0\rangle$ corresponds to doubly occupied
dot. The system of equations for these matrix elements has the form (\cite{Interplay})
\begin{eqnarray}
&&-\rho_0\left[D_+(\varepsilon_1)+D_+(\varepsilon_2)\right]+
\rho_1 [1-C_+(\varepsilon_1)]\nonumber\\
&&+\rho_2[1-C_+(\varepsilon_2)]- \rho_+
[C_-(\varepsilon_1)+C_-(\varepsilon_2)]=0,\label{17}\\
&&\rho_0 D_+(\varepsilon_1)-\rho_1 F_1+\rho_D
[1-C_+(\varepsilon_2+U)]\nonumber\\
&&+\rho_+[C_-(\varepsilon_1)+D_-(\varepsilon_2+U)]=0,\\
&&\rho_0 D_+(\varepsilon_2)-\rho_2 F_2+\rho_D
[1-C_+(\varepsilon_1+U)]\nonumber\\
&&+\rho_+[C_-(\varepsilon_2)+D_-(\varepsilon_1+U)]=0,\\
&&\rho_1 D_+(\varepsilon_2+U)+\rho_2
D_+(\varepsilon_1+U) \nonumber\\
&&-\rho_D[1-C_+(\varepsilon_1+U)]-\rho_D [1-C_+(\varepsilon_2+U)]\nonumber\\
&&-\rho_+[D_-(\varepsilon_1+U)+D_-(\varepsilon_2+U)]=0,\\
&&h\rho_+=\Gamma \rho_-F,\\
&&h\rho_-=-\Gamma(\rho_+F-\mathcal{L}/2).
\label{18}
\end{eqnarray}
In Eqs.~(\ref{17})-(\ref{18}) the following notations are used
\begin{eqnarray}
\rho_+=(\rho_{12}+\rho_{21})/2,\\
\rho_-=i(\rho_{12}-\rho_{21})/2,\\
C_\pm (\varepsilon)=\sum_n A_n [f_L(\varepsilon-n\omega)
\pm f_R(\varepsilon-n\omega)]/2,\label{19a}\\
D_\pm (\varepsilon)=\sum_n A_n [f_L(\varepsilon+n\omega)
\pm f_R(\varepsilon+n\omega)]/2,\\
\mathcal{L}=\rho_0[D_-(\varepsilon_1)+D_-(\varepsilon_2)]
\nonumber\\
+\rho_1[C_-(\varepsilon_1)+D_-(\varepsilon_2+U)]\nonumber \\
+\rho_2[C_-(\varepsilon_2)+D_-(\varepsilon_1+U)]
\nonumber\\
+\rho_D[C_-(\varepsilon_1+U)+C_-(\varepsilon_2+U)],\label{19c}\\
F_{1,2}=1-C_+(\varepsilon_{1,2})+D_+(\varepsilon_{2,1}+U),\label{19d}
\\
F=(F_1+F_2)/2,\label{19b}
\end{eqnarray}
where $f_\alpha(\varepsilon)=(\exp((\varepsilon-\mu_\alpha)/T_{\alpha})+1)^{-1}$
is the Fermi-Dirac distribution function 
and
\begin{equation}
A_n=\text{e}^{-\lambda^2(1+2n_B)}I_n(z)\text{e}^{n\hbar\omega/2T}
\end{equation}
with $I_n(z)$ is the
modified Bessel function of the first kind, $z=2\lambda^2\sqrt{n_B(1+n_B)}$, $n_B$ is the 
Bose-Einstein distribution function 
$n_B=(\exp(\hbar\omega/T)-1)^{-1}$. The temperature in the molecule is taken to be equal to the average temperature of the transistor, $T=(T_L+T_R)/2$.
Furthermore, the system of Eqs.~(\ref{17})-(\ref{18}) is complimented by normalization condition, $\rho_0+\rho_1+\rho_2+\rho_D=1$. 

In the case of zero bias voltage, $V=0$, the heat current is treated in this work via symmetrized expression, $J=(J_L+J_R)/2$. Then, the expression for the heat current in terms of the matrix elements of the density matrix of electrons in the dot acquires the following form 
\begin{eqnarray}\label{J}
&&J=\dfrac{\Gamma}{2\hbar}\left\{[(\varepsilon_1-\varepsilon_F)D_{-}(\varepsilon_1)+(\varepsilon_2-\varepsilon_F)D_{-}(\varepsilon_2)]\rho_0\right.\nonumber\\
&&+[(\varepsilon_1-\varepsilon_F)C_{-}(\varepsilon_1)+(\varepsilon_2-\varepsilon_F+U)D_{-}(\varepsilon_2+U)]\rho_1\nonumber\\
&&+[(\varepsilon_2-\varepsilon_F)C_{-}(\varepsilon_2)+(\varepsilon_1-\varepsilon_F+U)D_{-}(\varepsilon_1+U)]\rho_2\nonumber\\
&&+[(\varepsilon_1-\varepsilon_F+U)C_{-}(\varepsilon_1+U)\nonumber\\
&&+(\varepsilon_2-\varepsilon_F+U)C_{-}(\varepsilon_2+U)]\rho_D\\
&&-[(\varepsilon_1-\varepsilon_F)(1-C_{+}(\varepsilon_1))+(\varepsilon_1-\varepsilon_F+U)D_{+}(\varepsilon_1+U)\nonumber\\
&&\left.+(\varepsilon_2-\varepsilon_F)(1-C_{+}(\varepsilon_2))+(\varepsilon_2-\varepsilon_F+U)D_{+}(\varepsilon_2+U)]\rho_+\right\}\nonumber.
\end{eqnarray}

This expression, in couple with the system Eqs. (\ref{17})-(\ref{18}), is subjected to analysis by numerical methods.

\subsection{Heat current and negative differential thermoconductance.}

\begin{figure}\centering
\includegraphics[scale=0.5]{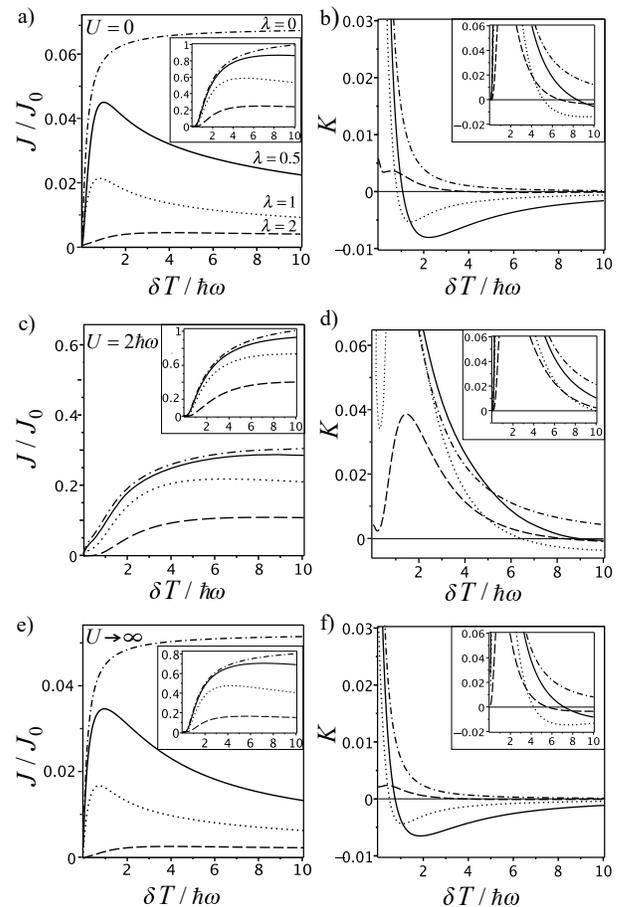}
\caption{\textit{Left panels}: The dependences of heat current $J$ in spintronic molecular transistor, normalized to $J_0=\hbar \omega I_0/e\;( I_0=e\Gamma/(2 \hbar)$, $\hbar \omega$ is the oscillation quantum), on temperature difference $\delta T$ at different values of electron-vibron interaction constant $\lambda$ and different Coulomb interaction constant (indicated in the figure). For non-zero electron-vibron interaction at intermediate temperatures $\delta T \simeq \hbar \omega $ negative differential thermoconductance occurs. \textit{Right panels}: The dependences of differential thermoconductance $K=\partial J/\partial \delta T$, normalized to $k_B\Gamma/(2\hbar)$, on temperature difference $\delta T$. The effect of negative differential thermoconductance is mostly pronounced for $0<\lambda\lesssim 1$. \textit{Parameters}: here and trough all the calculations in this paper the energy level width $\Gamma=0.0075\hbar\omega\ll T_R, T_L$; furthermore, $T_R=0.065\hbar\omega$ here and below, excluding Fig.~\ref{ris:P(T)}. Other used parameters:  $h=0.1 \hbar\omega$, combinations of electron-vibron interaction constant $\lambda$ and the detuning energy $\Delta$ (equals $\varepsilon_0-\varepsilon_F$, normalized to $\hbar \omega$), which correspond to similar regime of the device for each $\lambda$ read: (a)-(f), ($\lambda$; $\Delta$)$\rightarrow$ $(0;0.15) $ (dash-dot), $(0.5;0.4)$ (solid), $(1;1.15)$ (dot), $(2;4.15)$ (dash) -- small vicinity, much smaller than $\hbar \omega$, of resonance point;  insets in (a)-(f), $(0;2.15) $ (dash-dot), $(0.5;2.4)$ (solid), $(1;3.15)$ (dot), $(2;6.15)$ -- far distance, larger than $\hbar \omega$, from resonance point. The insets on (c),(d) show that the effect of negative differential thermoconductance is less pronounced for system with finite Coulomb interaction due to the superposition of contributions of elastic channels.}
\label{ris:J}
\end{figure}

With the help of the Eq.~(\ref{J}) the dependences of heat current and differential thermoconductance on temperature difference $\delta T$ have been numerically obtained within the restriction of applied approximation, that energy level width $\Gamma \ll T_R$, Fig.~\ref{ris:J}. 

The graphs in Figs.~\ref{ris:J} (a),(c),(e) show that for non-zero electron-vibron interaction, the heat current in the area of intermediate temperature differences ($\delta T \simeq \hbar \omega$) decreases with the increase of the temperature difference of the electrodes, i.e. the system is characterized by negative differential thermoconductance $K=\partial J/\partial \delta T$. Different panels represent the system for different values of Coulomb interaction $U$. The differential thermoconductance $K$ for corresponded set of parameters is plotted in Fig.~\ref{ris:J} (b),(d),(f), and the same holds for insets. One can see that differential thermal conductance takes on negative values for non-zero $\lambda$, having the most pronounced decay for small and intermediate coupling, $0<\lambda\lesssim 1$.

The effect of negative differential thermal conductance is caused by the vibronic effects. It is known that as long as the temperature of the system tends to zero the probability of existence of the vibron in the molecule is very small. So the probability of tunneling of electron with absorption of vibron is negligible. However, increasing of the temperature leads to significant changes caused by vibronic subsystem. For example, as it is shown in \cite{Monthioux},\cite{Ferone2},\cite{Interplay}, the electric conductance at strong electron-vibron interaction at intermediate temperatures has anomalous non-monotonic behaviour, due to the growth of the probability of tunneling caused by the opening of the channels with absorption of vibrons owing to the increase of temperature. 

Such changes of transport properties determine the non-monotonic behaviour of the heat current with a non-zero electron-vibron interaction, obtained and studied in this work on the example of spintronic molecular transistor.~$J$ grows slightly at $T\simeq \hbar\omega$ due to thermal enhancing. The suppression of heat current at bigger temperatures takes place due to polaronic effects, that generally affect the system under given parameter of electron-vibron interaction. Physically, molecule vibrations causes dissipation of energy during heat transmission process.

The effect takes place in a wide range of the detuning of the energy level in the molecule from the Fermi energy, $\Delta=\varepsilon_0-\varepsilon_F$, and most pronounced near the values corresponding to the resonant transmission of electrons through the system. Basing on Eqs.~(\ref{H_dot})-(\ref{E_s}) for the resonance points one finds: $ \Delta_{res\;*,1}= \lambda ^ 2 \mp  h / 2 $ and $ \Delta_{res\;2,**}= \lambda ^ 2 - U \mp h / 2 $. The curves in Fig.~\ref{ris:J} are calculated for optimal $\Delta$ near the resonance point, $ \Delta_ {res\;1} +\xi $, with $\xi\approx 0.1 \hbar \omega$. It will be clear in the following sections, that resonance points $ \Delta_ {res\;1} $ and $ \Delta_ {res\;2}$ are most important for the system, for the thermopower $S$ and thermoelectric power $P_{me}$ (calculated in linear response over $\delta T$), thus, one of them is chosen. Note that to analyze the location of the resonance points on the energy scale for several pairs of $\lambda$ and $U$, one can refer to the graphs of $ S (\Delta) $ in Fig.~\ref{ris:S}.

Each curve in insets in Fig.~\ref{ris:J} is calculated for $\Delta_ {res\;1} +2 \hbar \omega$ and shows that with a deviation $\simeq\hbar\omega$ far from the resonance point the decreasing of heat current with an increase of $\delta T$ is shifted to larger temperature differences and less pronounced. 

The Figs.~\ref{ris:J} (c),(d),(e),(f) demonstrate the influence of Coulomb interaction on thermocurrent and differential thermoconductance in spintronic molecular transistor. If the Coulomb interaction is non-zero, but finite, the effect of differential thermoconductance became less pronounced (and, consequently, it almost vanishes for $\Delta$, positioned in distance $\simeq\hbar\omega$ far from resonance point, Figs.~\ref{ris:J} (c), (d) insets). This is due to the superposition of contributions of elastic channels of the tunneling, when the quantum dot is singly occupied, and when it is doubly occupied, while the latter is thermally activated. Great energy contribution affects heat current. Due to that superposition, an increase of the heat current is maintained for more bigger values of $\delta T$. Moreover, when the slight decrease of its value with an increase of the temperature difference occurs, the system at this temperature differences will not be characterized by a significant value of the modulus of negative differential thermoconductance. At the same time, the system for $U = 0$ demonstrate the behaviour of $J$, while two coincided elastic tunneling channels are open. For $U\rightarrow \infty$ only one elastic channel is achievable. For the same reasons maximal heat current in this case is less than for $U=0$.

Heat current in spintronic molecular transistor demonstrates interesting non-monotonic behaviour caused by electron-vibron interaction and affected by Coulomb correlations.

\subsection{Thermopower in spintronic molecular transistor}

Let us move on to the next question: thermoelectric phenomena due to charge transfer under the applied temperature gradient. The problem is considered in linear response over bias voltage $eV\ll\varepsilon_F$ and temperature difference $\delta T\ll T$. In this case following relation for charge current $I$ is fulfilled
\begin{equation}\label{12}
I=GV+L\delta T.
\end{equation}
Here $G$ is the electric differential conductance in linear response, $L$ is the thermoelectric coefficient.  
For heat current in linear response one obtains
\begin{equation}
J=\tilde{L}V+K \delta T,
\end{equation}
where $\tilde{L}(-H)=TL(H)$ correspondingly to Casimir-Onsager relations, and $K$ is the differential thermal conductance. 

In open circuit in presence of temperature difference $\delta T$ between the two sides of the contact a voltage drop $V$ across the contact appears (Seebeck effect). If the contact is closed to an electron source, the bias voltage will cause the current to compensate the current due to the temperature difference. Correspondingly to the standard condition of compensation, i.e. zero total current, thermopower or Seebeck coefficient, which is defined as 
\begin{equation}
S=\left.\dfrac{V}{\delta T}\right|_{I=0,}
\end{equation}
in linear response reads
\begin{equation}\label{S}
S=-\dfrac{L}{G}.
\end{equation}
From a technological point of view this effect is vary important because it can be used to convert dissipated heat into electricity. This is also of great fundamental scientific interest because, by combining energy and charge flux data, it is possible to decode information about the dynamics of the system that is not available in charge transfer experiments alone(\cite{Di_Ventra}). Moreover, $S$ is used for calculation of the "quality factor" $ZT$ of thermoelectric machine, efficiency, thermoelectric power, etc. 

The thermopower has been calculated by direct differentiation 
\begin{equation}\label{S_partial}
S=-\dfrac{\left.(\partial I/\partial \delta T)\right|_{V\rightarrow 0, \delta T \rightarrow 0}}{\left.(\partial I/\partial V)\right|_{V\rightarrow 0, \delta T \rightarrow 0}},
\end{equation}
where the charge current, according to the previous calculations for this model \cite{Interplay} is
\begin{equation}
\frac{I}{I_0}=\frac{h^2}{h^2+\Gamma^2F^2}\;\mathcal{L},
\end{equation}
and $I_0= e\Gamma/(2\hbar)$, $e<0$ is the elementary charge, other notations are given in Eqs.~(\ref{19a})-(\ref{19b}).

It is known that in QD systems the thermopower is remarkable because it is possible to influence its sign and magnitude, particularly, by means of gate voltage (\cite{Di_Ventra},\cite{KochThermo}). Thermoelectric coefficient $L$, that enters Eq.~(\ref{S}), depends on the detuning energy $\Delta$ and governs the behaviour of the thermopower.

\begin{figure}\centering
\includegraphics[scale=0.5]{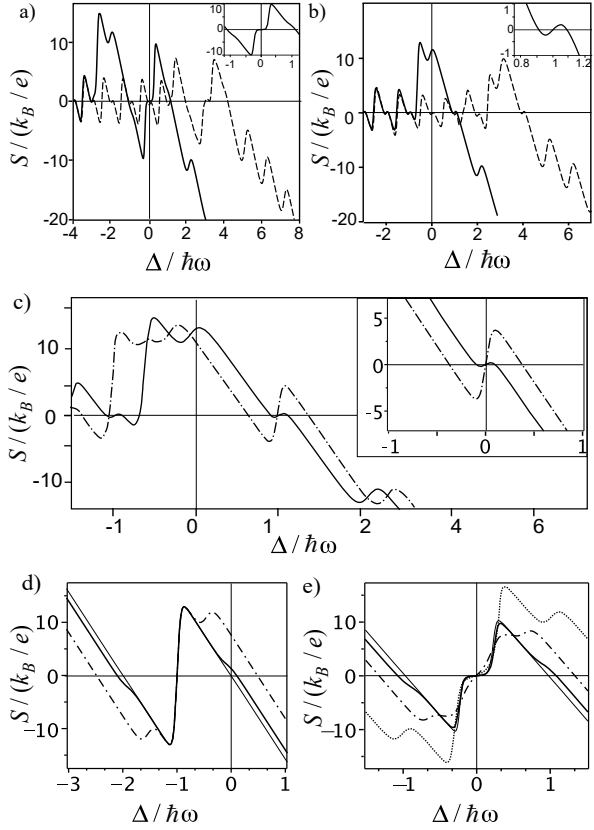}
\caption{Thermopower $S/(k_B/e)$ of spintronic molecular transistor depending on the detuning energy, $\Delta=\varepsilon_0-\varepsilon_F$. \textit{Upper panels}: the curves for different values of electron-vibron interaction constant $\lambda=1$ (solid), $2$ (dashed); a) $U=2\hbar \omega$. The emergence of addition points of function's inflection near the point of electron-hole symmetry of the system for non-zero Coulomb interaction; b) $U=0$. The emergence of addition points of changing of function's sign in the region between the resonance points due to an external magnetic field. Both dependences also show the sawtooth behaviour of the thermopower, which is common for molecular junctions. An external magnetic field $ h=0.25 \hbar \omega$; \textit{Middle and bottom panels} complement upper panels with the calculations for different values of an external magnetic field $h$: c) the dependences $S(\Delta)$ for $U=0$, $\lambda=1$, $h=0.25 \hbar \omega$ (solid), $0.75 \hbar \omega$ (dash-dotted); c) inset: curves for the same values of $h$ and $U$ in absence of electron-vibron interaction, $\lambda=0$; d) the dependences in absence of electron-vibron coupling, $\lambda=0$, for non-zero Coulomb interaction, $U=2\hbar \omega$, $h=0.1 \hbar \omega$ (solid thin), $0.25 \hbar \omega$ (solid thick), $0.75 \hbar \omega$ (dash-dotted); e) $U=2\hbar \omega$, electron-vibron coupling $\lambda=1$, $h=0.1\hbar \omega$ (solid thin), $0.25\hbar \omega$ (solid thick), $0.75\hbar \omega$ (dash-dotted), $2 \hbar \omega$ (dotted). Panels (a),(d) and (e) show that the characteristic "large step" of $S(\Delta)$ reveals the symmetry point. The curves depict, that in spintronic molecular transistor for non-zero Coulomb interaction there is a region of $\Delta$-s between the symmetry point and the resonance points, that yield non-zero thremopower in a wide range of magnetic fields.}
\label{ris:S}
\end{figure}

The expression for the thermopower $S$, Eq.~(\ref{S_partial}), has been treated numerically. The dependences of thermopower on the detuning energy are calculated, Fig.~\ref{ris:S}. It is found in the work that for $S(\Delta)$ the number of points of function's inflection and function's sign change increases in comparison with conventional molecular transistor. The feature is related to the presence of an external magnetic field and vibronic channels.

To discuss the effect, let us consider the system with Coulomb interaction. In nano-sized systems the detuning energy dependence of thermopower $S(\Delta)$ changes sign at resonance transmission points, where the energy level of the corresponding tunneling channel is equal to the Fermi level (and then the thermopower is zero by definition). And also changes sign at the point of electron-hole symmetry of the system, where the position of energy level in the dot relative to the Fermi energy is that the contribution to the current made by electrons is compensated by the contribution made by holes (\cite{Staring1},\cite{Swirkowicz},\cite{KochThermo},\cite{Zianni},\cite{Perroni}). It is known that resonance points in the presence of only elastic channels and also in the absence of an external magnetic field are the points $\Delta=0$ and $\Delta=-U$ (tunneling channels with single occupation of the dot and with double occupation respectively), and the point of electron-hole symmetry is $\Delta=-U/2$. Thus, a "large step" of thermopower, which is characteristic for nanosystems with Coulomb correlations of carriers (\cite{KochThermo},\cite{Swirkowicz},\cite{Perroni}), is formed. 

Fig.~\ref{ris:S} (a) presents the curves $S(\Delta)$ for non-zero Coulomb interaction ($U=2\hbar \omega$) for different values of electron-vibron interaction constant, $\lambda=1$ (solid line) and $2$ (dashed line). In Fig.~\ref{ris:S} (b) the parameters are the same, exept $U=0$. Furthermore, Figs.~\ref{ris:S} (a),(e) depict the emergence of addition points of function's inflection near the point of electron-hole symmetry. Figs.~\ref{ris:S} (b),(c) show the emergence of addition points of changing of function's sign in the region between the resonance points due to an external magnetic field. Figs.~\ref{ris:S} (c)-(e) present calculations for different values of magnetic field. The "large step" of $S(\Delta)$ in spintronic molecular transistor with $\lambda=1$ is shown in Figs.~\ref{ris:S} (a),(e), as well as in Fig.~\ref{ris:S} (d) for spintronic transistor with neglected electron-vibron interaction, $\lambda=0$. The symmetry point is $\Delta_{sym}=\lambda^2-U/2$. 

To describe the effects step by step, it is useful to consider the system with zero Coulomb interaction ($U = 0$) (Figs.~\ref{ris:S} (b),(c),(c) inset) and in the absence of electron-vibron interaction ($\lambda=0$) (Figs.~\ref{ris:S} (d),(c) inset) for comparison.  

Due to an external magnetic field the system has elastic channels corresponded to two energy levels $\varepsilon_1$ and $\varepsilon_2$, Eq.~(\ref{E_s}). Figs.~\ref{ris:S} (b),(c),(c) inset show that for $U=0$ the thermopower dependence on energy changes the sign at the points of resonance transmission, $\Delta_{res\,1,\,2}$, as it should be (note that for $U=0\;\Delta_{res\,2}\equiv\Delta_{res\,*}$). The number of resonance points increased (doubled) in comparison with conventional molecular transistor or the QD-based transistor without an applied external magnetic field and for negligible Coulomb interaction. Thus, the dependences acquire additional points of sign reversal. 

The graphs in Fig.~\ref{ris:S} (d) demonstrate that in the absence of electron-vibron coupling for non-zero Coulomb interaction the characteristic "large step" of thermopower reveals the symmetry point of the system.  Note that for $U\rightarrow \infty$ the position of the symmetry point formally moves to $-\infty$ and the behaviour of $S(\Delta)$ (near the resonance point $\Delta_{res\;1}$) reduplicate the behaviour of depicted curves. The effect of an external magnetic field for $\lambda=0$ is reduced to a shift of the resonance points and a shift of local peaks of thermopower.

However, it is manifested more significantly in the presence of electron-vibron interaction. Fig.~\ref{ris:S} (e) (and Fig.~\ref{ris:S} (a), which also depicts solid curve in another scale) shows that, for $U \ne 0,\,\lambda\ne0$  the thermopower near the symmetry point is affected by contributions due to the tunneling through inelastic channels (compare with the behaviour near the symmetry point in Fig.~\ref{ris:S} (d)). These channels are based on elastic channels, which number has increased due to Zeeman splitting. This changes the monotonicity of the thermopower in the symmetry point and $S(\Delta)$ acquires additional points of function's inflection. 

The magnitude of thermopower depends on the value of an external magnetic field. Furthermore, Fig.~\ref{ris:S} (e) shows that in spintronic molecular transistor there is a region of $\Delta$-s, that yield non-zero thermopower in a wide range of magnetic fields ($\Delta$-s between the symmetry point and resonance points). It is also holds that the position of the resonance points $\Delta_{res\,1,\,2}$ is shifted with increasing of the external magnetic field. Note that $\Delta_{res\,*,\,**}$ slightly affect the behaviour of the curve, but (due to temperature smoothing) do not allow $S$ to change sign.

The appearance of sawtooth behaviour of $S(\Delta)$ with steps of the order of a quantum of the molecule's oscillations (Figs.~\ref{ris:S} (a)-(c)) is caused by the electron-vibron interaction. The steps correspond to the opening of inelastic tunneling channels \cite{KochThermo},\cite{Zianni}.

Let us estimate the value of the thermopower reached near the resonance point (in Fig.~\ref{ris:S} (a), solid curve, $\Delta=1.15$). One gets $S\approx -1 k_B/e \approx -86*10^{-6}$V/K, that coincides by order of magnitude with the data given in the experiment \cite{Segalman}.

\subsection{Figure of merit $ZT$ and power of spintronic thermoelectric device.}
The optimal parameters of thermo machine have been found for non-zero Coulomb interaction and taking into account the electron-vibron interaction. Since the method is used within the perturbation theory, the range of temperatures $T\gg\Gamma$ is considered, which is, however, relevant for many experimentally realized tunnel systems. According to estimation based on the frequency of molecule oscillations from the experiment Ref.~\cite{ParkNature}, $\omega\approx 10^{12}$ Hz, the temperatures under consideration may be $\gtrsim 3$ K. Although, as it is known for transport through quantum dots and molecules, at lower temperatures the thermoelectric power can be even higher.
 
Consider the quantities of "quality factor" $ZT$ (figure of merit)
\begin{equation}
ZT=\dfrac{GS^2T}{K_e+K_{ph}},
\end{equation}
with $K_e=K-GS^2T$ being the electron contribution of $K$ to $ZT$ and $K_{ph}$ -- the phonon contribution, which is assumed to be negligibly small due to the mismatch at the interface "lead-dot", known in QD-systems (\cite{Murphy},\cite{Zubov});
maximal efficiency $\eta_m$ 
\begin{equation}
\eta_m=\eta_C\dfrac{\sqrt{ZT+1}-1}{\sqrt{ZT+1}+1},
\end{equation}
where $\eta_C=\delta T/T$ is the Carnot efficiency in linear response; and thermoelectric power at maximal efficiency $P_{me}$ 
\begin{equation}
P_{me}=\eta_C^2\dfrac{TK}{\sqrt{ZT+1}}\dfrac{\sqrt{ZT+1}-1}{\sqrt{ZT+1}+1}.
\end{equation}

Optimal parameters for considered spintronic thermoelectric device in the mentioned temperature range have been found. They correspond to a good "quality factor" $ZT$, which gives optimally large values of the efficiency $\eta_m$ and the highest power at maximum efficiency $P_{me}$. It has been found that maximal $P_{me}$ correspond to the values of $\Delta$ in the vicinity of $\Delta_{res\;1\,,2}$.  Let us fix one optimal value of the detuning energy $\Delta=\Delta_{res\;1} + \xi $ for each set of parameters. The second resonance point is equivalent in this sense, giving $\Delta=\Delta_{res\;2} -\xi $, however, for definiteness, that region is omitted. Optimal $P_{me}$ corresponds to $\Delta$ with $\xi\ll\hbar \omega $. The temperature smoothing  affects the parameters. For small temperatures, within the restriction where the method is applicable, it has been numerically found that $\xi\approx 0.1 \hbar \omega$. Let us analyse obtained dependences of the figure of merit $ZT$ and optimal power $P_{me}$ on temperature, external magnetic field, etc.

\begin{figure}\centering
\includegraphics[scale=0.7]{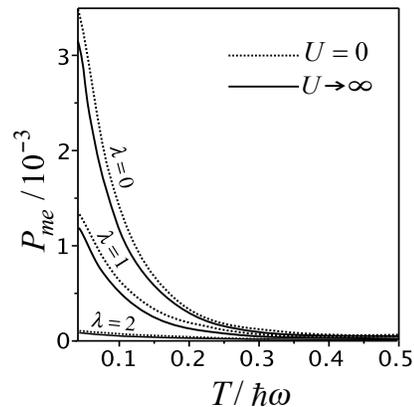}
\caption{The dependences of power at maximal efficiency $P_{me}/[(\delta T\,k_B)^2/\hbar]$ on temperature of the system $T$ for different values of electron-vibron interaction $\lambda$, indicated in the figure. The power at non-zero electron-vibron coupling is strongly suppressed comparing to the curves in non-interacting case. The effect of Coulomb interaction is also depicted: dotted curves correspond to the case of negligible Coulomb interaction, $U=0$, solid curves -- to $U\rightarrow \infty$; Here $h=0.1\hbar\omega$, the combinations ($\lambda;\Delta$) are the same as in main Fig.~\ref{ris:J} (not insets).}
\label{ris:P(T)}
\end{figure}

The dependence of the power at maximum efficiency $P_{me}$ on the temperature of the system is shown in Fig.~\ref{ris:P(T)}. The curves are calculated for optimal magnetic field $h=~0.1 \hbar\omega$ for different values of the electron-vibron interaction constant $\lambda$. One can conclude, that for considered ratio $T\gg\Gamma$ the electron-vibron coupling suppresses the thermoelectric power $P_{me}$ of the system. Dotted lines and solid lines represent the regimes with $U=0$ and $U\rightarrow \infty$ respectively.

\begin{figure}\centering
\includegraphics[scale=0.5]{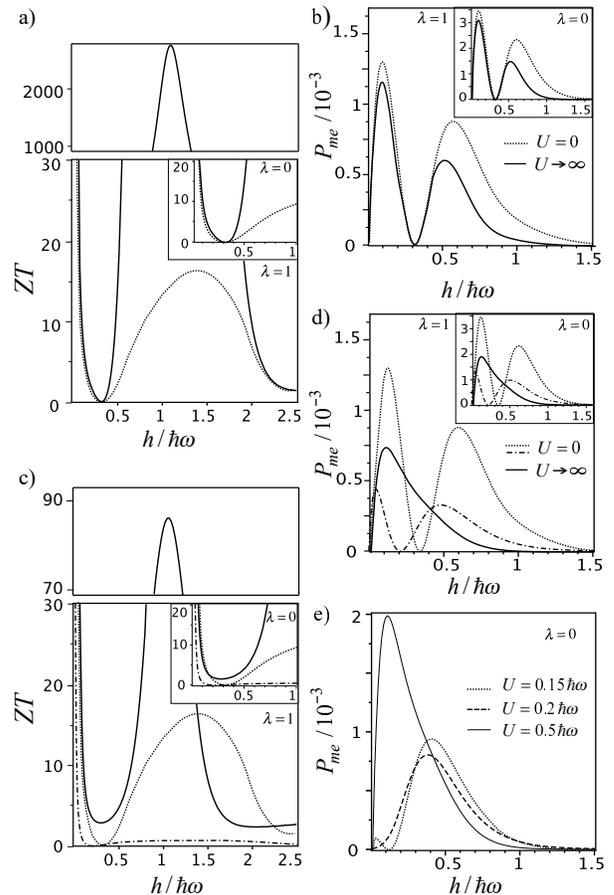}
\caption{ a) The dependences of figure of merit $ZT$ on an external magnetic field $h$ for electron-electron interaction $U=0$ (dotted) and $U \rightarrow \infty$ (solid) for spintronic molecular transistor for electron-vibron coupling constant $\lambda=1$. The combinations ($\lambda;\Delta$) are the same as in main Fig.~\ref{ris:J}. Inset (here and all over the figure) shows the dependences for the same parameters in the case $\lambda=0$. b) The dependences of power at maximal efficiency $P_{me}/[(\delta T\,k_B)^2/\hbar]$ on external magnetic field $h$ for the same parameters as the panel (a). For the optimal detuning energy $\Delta$ near the resonance point both curves $P_{me}(h)$ have two peaks and a local minimum; c) Figure of merit $ZT(h)$. The graphs differ from panel (a) by the parameter of the detuning energy. For $U\rightarrow \infty$ the detuning energy is set closer to the point of the electron-hole symmetry of the system, ($\lambda;\Delta$) is $(1;0.88)$ (solid). The curve has no points $ZT=0$. For $U=0$ it is set $(1;1.045)$ (dash-dotted), and $(1;1.16)$ (dotted), see the text, Sec. II C. For $U=0$ the behaviour of $ZT (h)$ is not drastically changed; c) inset: for $U\rightarrow \infty$ ($\lambda;\Delta$) is $(0;-0.12)$ (solid), for $U=0$ -- $(0;0.045)$ (dash-dotted), $(0; 0.16)$ (dotted); d) Curves $P_{me}(h)$ at the same parameters as the panel (c). The curve for $U\rightarrow \infty$ (solid) changes its behaviour while the detuning energy is closer to the point of the electron-hole symmetry. The power reaches its highest values in a wide \textit{continuous} range of external magnetic fields. Such a regime can be more handy for practical purposes. The curves for $U=0$ (dotted and dash-dotted), despite specific $\Delta$, do not drastically change theirs behaviour; e) $P_{me}(h)$ in transition region of $U$. $P_{me}(h)$ still has two maxima for $U=0.15\hbar\omega$ ($\Delta=-0.02$). The dependence became the curve with one peak for $U \gtrsim 0.2\hbar\omega$ ($\Delta=-0.045$). For $U \gtrsim 0.5\hbar\omega$ ($\Delta=-0.12$) the magnitude of $P_{me}$ became the same as for $U\rightarrow \infty$.}
\label{ris:P(h)}
\end{figure}

The panels in Fig.~\ref{ris:P(h)} illustrate the regions of the optimal external magnetic field $h$ for the system with non-zero electron-vibron interaction $\lambda$ for different values of the Coulomb interaction $U$: dotted and dash-dotted curves correspond to the system with $U=0$, solid curves -- to $U\rightarrow \infty$. In the Fig.~\ref{ris:P(h)} (e) $U$ is non-zero and $\lambda=0$, as well as in insets. 

Figs.~\ref{ris:P(h)} (a),(c) illustrate the behaviour of the figure of merit $ZT(h)$ for the same parameters, as the power $P_{me}(h)$ in Figs.~\ref{ris:P(h)} (b),(d) correspondingly. The parameters in (a) differ from ones in (c) by the value of detuning energy $\Delta$. Insets in Fig.~\ref{ris:P(h)} show the magnitude of the power $P_{me}$ and $ZT$ and the behaviour of the curves in the absence of electron-vibron interaction ($\lambda=0$).

The curves in Figs.~\ref{ris:P(h)} (a),(b) are calculated for optimal detuning energy near the resonance point, $\Delta_{res\,1}+\xi$. One can see from Fig.~\ref{ris:P(h)} (b), that for optimal detuning energy the dependences of the power $P_{me}$ on an external magnetic field $h$ in spintronic molecular transistor have two peaks and a local minimum associated with the vanishing of the thermopower $S$, and, hence, the figure of merit $ZT$ (for reference in the case $\lambda=0$ see also \cite{Zubov}). However, Figs.~\ref{ris:S} (a),(d),(e) in previous subsection depict, that in the case of a non-zero Coulomb interaction there is a "large step" of the dependence of the thermopower $S(\Delta)$ on the detuning energy, that causes the thermopower not to vanish to zero with increasing magnetic field for $\Delta$ close to the point of electron-hole symmetry of the system. This allows one to expect, that for values of the detuning energy in the interval between the resonance and the symmetry points, the dependence of the power $P_{me}$ on an external magnetic field will have only one peak and after the growth $P_{me}(h)$ will be monotonic with increasing magnetic field.

In support of these qualitative conclusions, numerically calculated curves show, that for $U\rightarrow\infty$ for $\Delta$ closer to the symmetry point the figure of merit $ZT$ does not achieve zero value for any magnetic field (Fig.~\ref{ris:P(h)} (c), solid curve). Consequently, the power $P_{me}$ for the same parameters reaches its highest values in a wide \textit{continuous} range of external magnetic fields (Fig.~\ref{ris:P(h)} (d), solid curve). For calculation the $\Delta=\Delta_{res_{1}}+\xi^{\prime}$, with $\xi^{\prime}\approx -0.17$, the detuning energy of sub-peak of $P_{me}(\Delta)$ for optimal field $h=0.1 \hbar \omega$, were chosen. $\Delta$ should be any from interval between the resonance point and symmetry point, i.e. mentioned value is chosen for reasons of obtaining of high values of $P_{me}$.

On the other hand, for $U=0$ the effect does not take place (Fig.~\ref{ris:P(h)} (d), dash-dotted curve and dotted curve), even while the detuning energy corresponds to $\Delta$ from the region between the resonance points. For example, the dash-dotted curve is calculated for $\Delta$ belonging to an interval between the resonance points of $P_{me}(\Delta)$, while $h=0.1\hbar \omega$ (i.e. ($\lambda$; $\Delta$)$\rightarrow $ $(1;1.045)$ in Fig.~\ref{ris:P(h)} (d), and $(0;0.045)$ in Fig.~\ref{ris:P(h)} (d) inset). For calculation of the dotted curve the $\Delta$ is set to be corresponding to second maximum of $P_{me}(\Delta)$, appearing at strong fields, e.g. $h=0.75\hbar\omega$. The curves for $U=0$ with $(1;1.16)$ (Fig.~\ref{ris:P(h)} (d)) and $U=0$ with $(0;0.16) $ (Fig.~\ref{ris:P(h)} (d) inset) are given. 

The effect of appearance of continuous region of magnetic fields that corresponds to high values of thermoelectric power can be useful for the reasons of more handy application of thermoelectric device, integration of spintronic SMT into circuits, etc.

Furthermore, for too small $U\ll \hbar \omega$ the curve $P_{me}(h)$ remains qualitatively the same, as for $U=0$, see Fig.~\ref{ris:P(h)} (e)). Indeed, due to temperature smoothing the emergence of "large step"  for small $U$ is improbable.  The energy scale $|\xi^{\prime}|, \,\xi\ll \hbar \omega $ is appropriate as the distance of significant changes near the resonance point. It should be expected, that $U/2\gtrsim \xi$ or, more precisely, $|\xi^{\prime}|-h/2$  (with $h=0.1\hbar\omega$ correspondingly to calculation), may be enough Coulomb interaction, thus, the interval between the symmetry point and the resonance point will be discernible for the system. Indeed, in the considered range of parameters, $ T \gg \Gamma $, it has been obtained, that it is necessary to fulfil the condition $ U\gtrsim 0.2\hbar\omega$ to get the effect of changing of the form of $ P_{me}(h) $, i.e. to get special $ \Delta $-s, for which $ S(\Delta) $ does not turn into $ 0 $ with growing $ h $ (Fig.~\ref{ris:P(h)} (e), where the transition region of $ U $ is depicted for $ \lambda = 0 $). In Fig.~\ref{ris:P(h)} (e)  the dependence $P_{me}(h)$ for $U=0.15\hbar \omega$ still has two maxima despite specific $\Delta$ belonging to an interval between the resonance points.  Moreover, already for $ U \gtrsim 0.5 \hbar\omega$ the power reaches the same maximal values as for $U\rightarrow \infty$.

Note that, whenever for $ U \rightarrow0$ such $ \Delta $ that $ S (\Delta) $ is not equal to zero up to strong fields $ h \gg \hbar \omega $ can be found, this one will be positioned far apart from the resonance points. Then $ P_{me} \rightarrow 0 $, and this feature is not considered due to lack of practical interest. Note also that for $ U \rightarrow0$ for magnetic fields $ h\gg U $, the dependence $ P_{me} (\Delta) $ can have a second maximum of a different nature (see $S(\Delta)$ in Fig.~\ref{ris:S} (c), (c) inset in previous subsection, the second maximum of $P_{me}$ corresponds to the $\Delta$ belonging to an interval between the resonance points). Due to this qualitative distinction, the dependence $ P_{me}(h) $ for all $ \Delta $, as before, will have the same form as for $ U = 0 $ (Fig.~\ref{ris:P(h)} (d), dotted curve).

As it has been mentioned, far from the points of resonance transmission the magnitude of power will be small. This is due to the fact, that for high values of power $P_{me}$, finite but not too large values of $ZT$  are required, as it is well known (see, e. g. \cite{Zubov}, \cite{Khedri}) and obvious when matching the Figs.~\ref{ris:P(h)} (a) and (b), Figs.~\ref{ris:P(h)} (c) and (d). But $ZT$, as well as $S$ (Fig.~\ref{ris:S}), is larger far from resonance point. It should also be taken into account, that the power $P_{me}$ is governed by the magnitude of the differential thermoconductance in linear response and by other transport characteristics. For these reasons the features of the behaviour of the thermopower $S$ near the symmetry point for non-zero electron-vibron interaction, Figs.~\ref{ris:S} (a),(e), do not cause significant changes of $P_{me}$. Furthermore, as shown in Fig.~\ref{ris:P(h)} (a),(c), for non-zero electron-vibron interaction the figure of merit $ZT$ at strong magnetic fields is highly damped. Nevertheless, it does not affect $P_{me}(h)$ curves. 

Note that maximum magnitude of the power for $\Delta$ closer to the symmetry point is quite smaller than in optimal point (compare maximum values of solid lines in Figs.~\ref{ris:P(h)} (d) and \ref{ris:P(h)} (b)). Despite of slight suppressing of magnitude of $P_{me}$, the continuity of the region of magnetic fields, that corresponds to the highest values of power, can be a good characteristic for more handy using of thermoelectric device for practical purposes. 

Estimating the values of fields under consideration using the experimental $\hbar\omega$ \cite{ParkNature} one will have: $h$ from $0.03 \hbar\omega$ to $0.75 \hbar\omega$ is the range approximately from $1$ to $30$ T. Optimal field is $h=0.1\hbar\omega\approx 4$ T.

\section{Conclusions}
QD-based nanothermoelectric machines are promising as efficient converters of thermal energy into electrical energy due to high values of thermopower. Consequently, it is an important task to study heat transport properties and identify the conditions for optimal thermopower, $ZT$, thermoelectric power at maximal efficiency etc. in nanodevices. 

In present work the thermoelectric phenomena in molecular transistor with spin-polarized electrodes in an external magnetic field have been studied.

It has been obtained that the system is characterized by negative differential thermoconductance at non-zero electron-vibron interaction. Such a behaviour of differential thermal conductance reveals an existence of nonlinear processes in the molecule-based system and strong influence of the electron-vibron interaction on real device if the oscillations of the middle island cannot be neglected. 

It has been found that in spintronic molecular transistor the number of points of change of the sign and points of function's inflection of the dependence of thermopower on detuning energy increases affecting the magnitude of $S$ due to Zeeman splitting and vibronic effects.

Optimal parameters and regimes of the device, corresponded to best values of figure of merit $ZT$ and thermoelectric power at maximal efficiency $P_{me}$, have been obtained. The range of parameters corresponded to regime with continuous interval of magnetic fields with maximal thermoelectic power has been found. Such a regime can be useful for the reasons of more handy application of thermoelectric system. It differs from ordinary regime of magnetic device, where the dependence $P_{me}$ on magnetic field has two maxima and a local minimum. More handy regime is reached, when detuning energy is set to be closer to the point of electrone-hole symmetry of the system. The effect is caused by the appearance of the "large step" of thermopower dependence on the detuning energy.

Thus, in addition to controling by gate voltage, the effects make possible to achieve optimal thermoelectric characteristics and control thermoelectric properties of nanosize thermo machines by means of an external magnetic field. Furthermore, the results allows one to account for changes in thermoelectric characteristics in the case of non-zero electron-vibron coupling.


 \textbf{\emph{Acknowledgement}}

The author thanks I.V.~Krive for conceptualization of the problem and useful advices and S.I.~Kulinich, O.A.~Ilinskaya and O.M.~Bahrova for fruitful discussions. This work is supported by the National Academy of Sciences of Ukraine (Scientific Program 1.4.10.26.4).

\end{document}